*Article*

# Evo-SETI: A Mathematical Tool for Cladistics, Evolution, and SETI

**Claudio Maccone** [1,2]

[1]  International Academy of Astronautics (IAA) and IAA SETI Permanent Committee; IAA, 6 Rue Galilée, 75016 Paris, France; clmaccon@libero.it; Tel.: +39-342-354-3295

[2]  Istituto Nazionale di Astrofisica (INAF), Via Martorelli 43, 10155 Torino (TO), Italy

**Abstract:** The discovery of new exoplanets makes us wonder *where* each new exoplanet *stands* along its way to develop life as we know it on Earth. Our Evo-SETI Theory is a mathematical way to face this problem. We describe cladistics and evolution by virtue of a few statistical equations based on lognormal probability density functions (pdf) *in the time*. We call *b*-lognormal a lognormal pdf starting at instant *b* (birth). Then, the lifetime of any living being becomes a suitable *b*-lognormal *in the time*. Next, our *"Peak-Locus Theorem" translates cladistics*: each species created by evolution is a *b*-lognormal whose peak lies on the *exponentially growing* number of living species. This exponential is the *mean value of a stochastic process* called "Geometric Brownian Motion" (GBM). Past mass extinctions were all-lows of this GBM. In addition, the Shannon Entropy (with a reversed sign) of each *b*-lognormal is the measure of how evolved that species is, and we call it EvoEntropy. The "molecular clock" is re-interpreted as the EvoEntropy straight line in the time whenever the mean value is exactly the GBM exponential. We were also able to extend the Peak-Locus Theorem to any mean value other than the exponential. For example, we derive in this paper for the first time the EvoEntropy corresponding to the Markov-Korotayev (2007) "cubic" evolution: a curve of logarithmic increase.

---

## 1. Purpose of This Paper

This paper describes the recent developments in a new statistical theory describing Evolution and SETI by mathematical equations. I call this the Evo-SETI mathematical model of Evolution and SETI.

The main question which this paper focuses on is, whenever a new exoplanet is discovered, what is the evolutionary stage of the exoplanet in relation to the life on it, compared to how it is on Earth today? This is the central question for Evo-SETI. In this paper, it is also shown that the (Shannon) Entropy of *b*-lognormals addresses this question, thus allowing the creation of an Evo-SETI SCALE that may be applied to exoplanets.

An important new result presented in this paper stresses that the cubic in the work of Markov-Korotayev [1–8] can be taken as the mean value curve of a lognormal process, thus reconciling their deterministic work with our probabilistic Evo-SETI theory.

## 2. During the Last 3.5 Billion Years, Life Forms Increased as in a (Lognormal) Stochastic Process

Figure 1 shows the time *t* on the horizontal axis, with the convention that negative values of *t* are past times, zero is now, and positive values are future times. The starting point on the time axis is *ts* = 3.5 billion ($10^9$) years ago, i.e., the accepted time of the origin of life on Earth. If the origin of life started earlier than that, for example 3.8 billion years ago, the following equations would remain the same and their numerical values would only be slightly changed. On the vertical axis is the number of species living on Earth at time $t$, denoted $L(t)$ and standing for "life at time $t$". We do not know this "function of the time" in detail, and so it must be regarded as a random function, or stochastic process $L(t)$. This paper adopts the convention that capital letters represent random variables, i.e., stochastic processes if they depend on the time, while lower-case letters signify ordinary variables or functions.

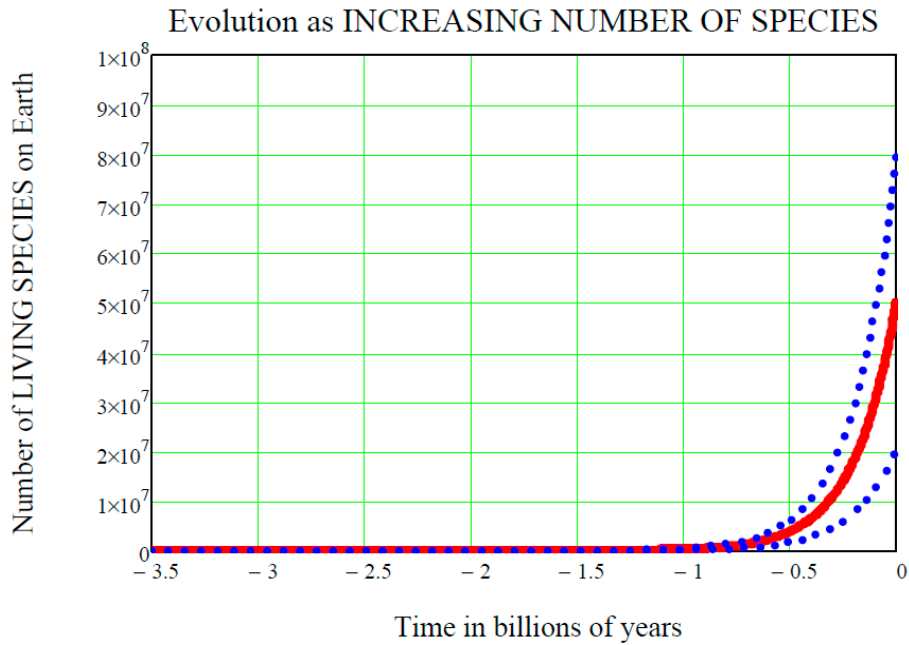

**Figure 1. Increasing DARWINIAN EVOLUTION as the increasing number of living species on Earth between 3.5 billion years ago and now.** The red solid curve is the mean value of the GBM stochastic process $L_{GBM}(t)$ given by Equation (22) (with $t$ replaced by ($t$-$ts$)), while the blue dot-dot curves above and below the mean value are the two standard deviation upper and lower curves, given by Equations (11) and (12), respectively, with $m_{GBM}(t)$ given by Equation (22). The "Cambrian Explosion" of life, that on Earth started around 542 million years ago, is evident in the above plot just before the value of −0.5 billion years in time, where all three curves "start leaving the time axis and climbing up". Notice also that the starting value of living species 3.5 billion years ago is *one* by definition, but it "looks like" zero in this plot since the vertical scale (which is the true scale here, not a log scale) does not show it. Notice finally that nowadays (i.e., at time $t = 0$) the two standard deviation curves have exactly the same distance from the middle mean value curve, i.e., 30 million living species more or less the mean value of 50 million species. These are assumed values that we used just to exemplify the GBM mathematics: biologists might assume other numeric values.

## 3. Mean Value of the Lognormal Process *L*(*t*)

The most important, ordinary and continuous function of the time associated with a stochastic process like $L(t)$ is its mean value, denoted by:

$$m_L(t) \equiv \langle L(t) \rangle. \tag{1}$$

The probability density function (*pdf*) of a stochastic process like $L(t)$ is assumed in the Evo-SETI theory to be a *b*-lognormal, and its equation thus reads:

$$L(t)\_pdf(n; M_L(t), \sigma, t) = \frac{e^{-\frac{[\ln(n) - M_L(t)]^2}{2\sigma_L^2(t-ts)}}}{\sqrt{2\pi}\,\sigma_L\sqrt{t-ts}\,n} \text{ with } \begin{cases} n \geq 0, \\ t \geq ts, \end{cases} \text{ and } \begin{cases} \sigma_L \geq 0, \\ M_L(t) = \text{arbitrary function of } t. \end{cases} \tag{2}$$

This assumption is in line with the extension in time of the statistical Drake equation, namely the foundational and statistical equation of SETI, as shown in [9].

The mean value (Equation (1)) is related to the pdf Equation (2) by the relevant integral in the number $n$ of living species on Earth at time $t$, as follows:

$$m_L(t) \equiv \int\limits_0^\infty n \cdot \frac{e^{-\frac{[\ln(n) - M_L(t)]^2}{2\,\sigma_L^2\,(t-ts)}}}{\sqrt{2\pi}\,\sigma_L \sqrt{t-ts}\,n}\, dn\,. \tag{3}$$

The "surprise" is that this integral in Equation (3) may be exactly computed with the key result, so that the mean value $m_L(t)$ is given by:

$$m_L(t) = e^{M_L(t)}\, e^{\frac{\sigma_L^2}{2}\,(t-ts)}\,. \tag{4}$$

In turn, the last equation has the "surprising" property that it may be exactly inverted, i.e., solved for $M_L(t)$:

$$M_L(t) = \ln(m_L(t)) - \frac{\sigma_L^2}{2}\,(t-ts)\,. \tag{5}$$

## 4. *L(t)* Initial Conditions at *ts*

In relation to the initial conditions of the stochastic process $L(t)$, namely concerning the value $L(ts)$, it is assumed that the exact positive number

$$L(ts) = Ns \tag{6}$$

is always known, i.e., with a probability of one:

$$\Pr\{L(ts) = Ns\} = 1\,. \tag{7}$$

In practice, $Ns$ will be equal to one in the theories of the evolution of life on Earth or on an exoplanet (i.e., there must have been a time $ts$ in the past when the number of living species was just one, be it RNA or something else), and it is considered as equal to the number of living species just before the asteroid/comet impacted in the theories of mass extinction of life on a planet.

The mean value $m_L(t)$ of $L(t)$ must also equal the initial number $Ns$ at the initial time $ts$, that is:

$$m_L(ts) = Ns\,. \tag{8}$$

Replacing $t$ with $ts$ in Equation (4), one then finds:

$$m_L(ts) = e^{M_L(ts)}\,. \tag{9}$$

That, checked against Equation (8), immediately yields:

$$Ns = e^{M_L(ts)} \text{ that is } M_L(ts) = \ln(Ns)\,. \tag{10}$$

These are the initial conditions for the mean value.

After the initial instant $ts$, the stochastic process $L(t)$ unfolds, oscillating above or below the mean value in an unpredictable way. Statistically speaking however, it is expected that $L(t)$ does not "depart too much" from $m_L(t)$, and this fact is graphically shown in Figure 1 by the two dot-dot blue curves above and below the mean value solid red curve $m_L(t)$. These two curves are the upper standard deviation curve

$$\text{upper\_st\_dev\_curve}(t) = m_L(t)\left[1 + \sqrt{e^{\sigma_L^2(t-ts)} - 1}\right] \tag{11}$$

and the lower standard deviation curve

$$\text{lower\_st\_dev\_curve}(t) = m_L(t) \left[ 1 - \sqrt{e^{\sigma_L^2(t-ts)} - 1} \right] \tag{12}$$

respectively (see [4]). Both Equations (11) and Equations (12), at the initial time $t = ts$, equal the mean value $m_L(ts) = Ns$. With a probability of one, the initial value $Ns$ is the same for all of the three curves shown in Figure 1. The function of the time

$$\text{variation\_coefficient}(t) = \sqrt{e^{\sigma_L^2(t-ts)} - 1} \tag{13}$$

is called the variation coefficient, since the standard deviation of $L(t)$ (noting that this is just the standard deviation $\Delta_L(t)$ of $L(t)$ and not either of the above two "upper" and "lower" standard deviation curves given by Equations (11) and (12), respectively) is:

$$\text{st\_dev\_curve}(t) \equiv \Delta_L(t) = m_L(t) \sqrt{e^{\sigma_L^2(t-ts)} - 1}. \tag{14}$$

Thus, Equation (14) shows that the variation coefficient of Equation (13) is the ratio of $\Delta_L(t)$ to $m_L(t)$, i.e., it expresses how much the standard deviation "varies" with respect to the mean value. Having understood this fact, the two curves of Equations (11) and (12) are obtained:

$$m_L(t) \pm \Delta_L(t) = m_L(t) \pm m_L(t) \sqrt{e^{\sigma_L^2(t-ts)} - 1}. \tag{15}$$

## 5. $L(t)$ Final Conditions at $te > ts$

With reference to the final conditions for the mean value curve, as well as for the two standard deviation curves, the final instant can be termed $te$, reflecting the end time of this mathematical analysis. In practice, this $te$ is zero (i.e., now) in the theories of the evolution of life on Earth or exoplanets, but it is the time when the mass extinction ends (and life starts to evolve again) in the theories of mass extinction of life on a planet. First of all, it is clear that, in full analogy to the initial condition Equation (8) for the mean value, the final condition has the form:

$$m_L(te) = Ne \tag{16}$$

where $Ne$ is a positive number denoting the number of species alive at the end time $te$. However, it is not known what random value $L(te)$ will take, but only that its standard deviation curve Equation (14) will, at time $te$, have a certain positive value that will differ by a certain amount $\delta Ne$ from the mean value Equation (16). In other words, from Equation (14):

$$\delta Ne = \Delta_L(te) = m_L(te) \sqrt{e^{\sigma_L^2(te-ts)} - 1}. \tag{17}$$

When dividing Equation (17) by Equation (16), the common factor $m(te)$ is cancelled out, and one is left with:

$$\frac{\delta Ne}{Ne} = \sqrt{e^{\sigma_L^2(te-ts)} - 1}. \tag{18}$$

Solving this for $\sigma_L$ finally yields:

$$\sigma_L = \frac{\sqrt{\ln \left[ 1 + \left( \frac{\delta Ne}{Ne} \right)^2 \right]}}{\sqrt{te - ts}}. \tag{19}$$

This equation expresses the so far unknown numerical parameter $\sigma_L$ in terms of the initial time $ts$ plus the three final-time parameters ($te$, $Ne$, $\delta Ne$).

Therefore, in conclusion, it is shown that once the five parameters ($ts$, $Ns$, $te$, $Ne$, $\delta Ne$) are assigned numerically, the lognormal stochastic process $L(t)$ is completely determined.

Finally, notice that the square of Equation (19) may be rewritten as:

$$\sigma_L^2 = \frac{\ln\left[1 + \left(\frac{\delta Ne}{Ne}\right)^2\right]}{te - ts} = \ln\left\{\left[1 + \left(\frac{\delta Ne}{Ne}\right)^2\right]^{\frac{1}{te-ts}}\right\} \tag{20}$$

from which the following formula is inferred:

$$e^{\sigma_L^2} = e^{\ln\left\{\left[1 + \left(\frac{\delta Ne}{Ne}\right)^2\right]^{\frac{1}{te-ts}}\right\}} = \left[1 + \left(\frac{\delta Ne}{Ne}\right)^2\right]^{\frac{1}{te-ts}}. \tag{21}$$

This Equation (21) enables one to get rid of $e^{\sigma_L^2}$, replacing it by virtue of the four boundary parameters: ($ts$, $te$, $Ne$, $\delta Ne$). It will be later used in Section 8 to rewrite the Peak-Locus Theorem in terms of the boundary conditions, rather than in terms of $e^{\sigma_L^2}$.

## 6. Important Special Cases of $m(t)$

(1)  The particular case of Equation (1) when the mean value $m(t)$ is given by the generic exponential:

$$m_{\text{GBM}}(t) = N_0\, e^{\mu_{\text{GBM}} t} = \text{or, alternatively,} = A\, e^{B\, t} \tag{22}$$

is called the Geometric Brownian Motion (GBM), and is widely used in financial mathematics, where it represents the "underlying process" of the stock values (Black-Sholes models). This author used the GBM in his previous models of Evolution and SETI ([9–14]), since it was assumed that the growth of the number of ET civilizations in the Galaxy, or, alternatively, the number of living species on Earth over the last 3.5 billion years, **grew exponentially** (Malthusian growth). Upon equating the two right-hand-sides of Equations (4) and (22) (with $t$ replaced by ($t$-$ts$)), we find:

$$e^{M_{\text{GBM}}(t)}\, e^{\frac{\sigma_{\text{GBM}}^2}{2}(t-ts)} = N_0\, e^{\mu_{\text{GBM}}(t-ts)}. \tag{23}$$

Solving this equation for $M_{\text{GBM}}(t)$ yields:

$$M_{\text{GBM}}(t) = \ln N_0 + \left(\mu_{\text{GBM}} - \frac{\sigma_{\text{GBM}}^2}{2}\right)(t - ts). \tag{24}$$

This is (with $ts = 0$) the mean value at the exponent of the well-known GBM pdf, i.e.,:

$$\text{GBM}(t)\_pdf(n; N_0, \mu, \sigma, t) = \frac{e^{-\frac{\left[\ln(n) - \left(\ln N_0 + \left(\mu - \frac{\sigma^2}{2}\right) t\right)\right]^2}{2\sigma^2 t}}}{\sqrt{2\pi}\, \sigma \sqrt{t}\, n}, \ (n \geq 0). \tag{25}$$

This short description of the GBM is concluded as the exponential sub-case of the general lognormal process Equation (2), by warning that GBM is a misleading name, since GBM is a lognormal process and not a Gaussian one, as the Brownian Motion is.

(2)　As has been mentioned already, another interesting case of the mean value function $m(t)$ in Equation (1) is when it equals a generic **polynomial in t starting at *ts***, namely (with $c_k$ being the coefficient of the $k$-th power of the time $t$-$ts$ in the polynomial)

$$m_{\text{polynomial}}(t) = \sum_{k=0}^{\text{polynomial\_degree}} c_k \, (t - ts)^k. \tag{26}$$

The case where Equation (26) is a second-degree polynomial (i.e., a parabola in $t - ts$) may be used to describe the Mass Extinctions on Earth over the last 3.5 billion years (see [13]).

(3)　Having so said, the notion of a *b*-lognormal must also be introduced, for $t > b =$ birth, representing the lifetime of living entities, as single cells, plants, animals, humans, civilizations of humans, or even extra-terrestrial (ET) civilizations (see [12], in particular pages 227–245)

$$\text{b} - \text{lognormal\_pdf}(t; \mu, \sigma, b) = \frac{e^{-\frac{[\ln (t-b) - \mu]^2}{2\sigma^2}}}{\sqrt{2\pi}\,\sigma\,(t - b)}. \tag{27}$$

## 7. Boundary Conditions when $m(t)$ is a First, Second, or Third Degree Polynomial in the Time ($t$-$ts$)

In [13], the reader may find a mathematical model of Darwinian Evolution different from the GBM model. That model is the Markov-Korotayev model, for which this author proved the mean value (1) to be a Cubic($t$) i.e., a third degree polynomial in $t - ts$.

In summary, the key formulae proven in [13], relating to the case when the assigned mean value $m_L(t)$ is a polynomial in $t$ starting at $ts$, can be shown as:

$$m_L(t) = \sum_{k=0}^{\text{polynomial\_degree}} c_k (t - ts)^k. \tag{28}$$

(1)　**The mean value is a straight line.** This straight line is the line through the two points, $(ts, Ns)$ and $(te, Ne)$, that, after a few rearrangements, becomes:

$$m_{\text{straight\_line}}(t) = (Ne - Ns)\frac{t - ts}{te - ts} + Ns. \tag{29}$$

(2)　**The mean value is a parabola**, i.e., a quadratic polynomial in $t - ts$. Then, the equation of such a parabola reads:

$$m_{\text{parabola}}(t) = (Ne - Ns)\frac{t - ts}{te - ts}\left[2 - \frac{t - ts}{te - ts}\right] + Ns. \tag{30}$$

Equation (30) was actually firstly derived by this author in [13] (pp. 299–301), in relation to Mass Extinctions, i.e., it is a decreasing function of time.

(3)　**The mean value is a cubic**. In [13] (pp. 304–307), this author proved, in relation to the Markov-Korotayev model of Evolution, that the ***cubic*** mean value of the $L(t)$ lognormal stochastic process is given by the cubic equation in $t - ts$:

$$m_{\text{cubic}}(t) = (Ne - Ns) \cdot \frac{(t - ts)\left[2(t - ts)^2 - 3(t_{\text{Max}} + t_{\text{min}} - 2\,ts)(t - ts) + 6(t_{\text{Max}} - ts)(t_{\text{min}} - ts)\right]}{(te - ts)\left[2(te - ts)^2 - 3(t_{\text{Max}} + t_{\text{min}} - 2\,ts)(te - ts) + 6(t_{\text{Max}} - ts)(t_{\text{min}} - ts)\right]} + Ns. \tag{31}$$

Notice that, in Equation (31), one has, in addition to the usual initial and final conditions $Ns = m_L(ts)$ and $Ne = m_L(te)$, two more "middle conditions" referring to the two instants ($t_{Max}$, $t_{min}$) at which the Maximum and the minimum of the cubic $Cubic(t)$ occur, respectively:

$$\begin{cases} t_{min} = \text{time\_of\_the\_Cubic\_minimum} \\ t_{Max} = \text{time\_of\_the\_Cubic\_Maximum.} \end{cases} \qquad (32)$$

## 8. Peak-Locus Theorem

The Peak-Locus theorem is the new mathematical discovery of ours, playing a central role in Evo-SETI. In its most general formulation, it can be used for any lognormal process $L(t)$ or arbitrary mean value $m_L(t)$. In the case of GBM, it is shown in Figure 2.

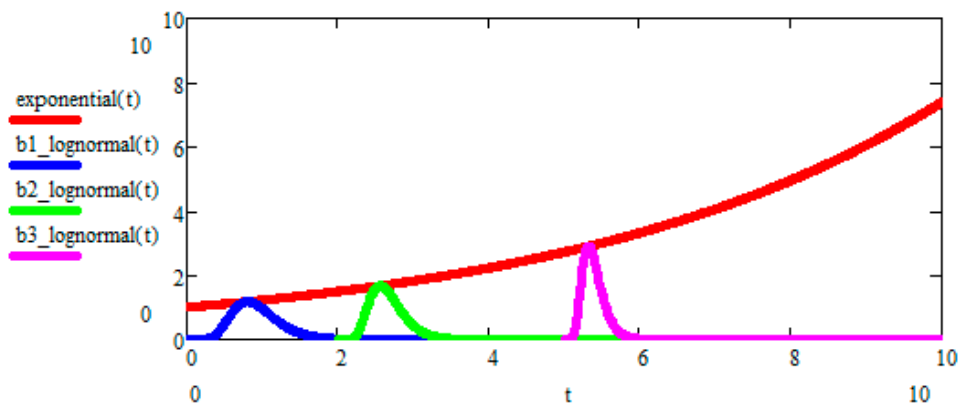

**Figure 2.** **The Darwinian Exponential is used as the geometric locus of the peaks of *b*-lognormals for the GBM case.** Each *b*-lognormal is a lognormal starting at a time *b* (birth time) and represents a different **species** that originated at time *b* of the Darwinian Evolution. This is **cladistics**, as seen from the perspective of the Evo-SETI model. It is evident that, when the generic "running *b*-lognormal" moves to the right, its peak becomes higher and narrower, since the area under the *b*-lognormal always equals one. Then, the (Shannon) **entropy** of the running *b*-lognormal is the **degree of evolution** reached by the corresponding **species** (or living being, or a civilization, or an ET civilization) in the course of Evolution (see, for instance, [14–19]).

The Peak-Locus theorem states that the family of *b*-lognormals, each having its own peak located exactly ***upon*** the mean value curve (1), is given by the following three equations, specifying the three parameters $\mu(p)$, $\sigma(p)$, and $b(p)$ appearing in Equation (27) as three functions of the peak abscissa, i.e., the independent variable *p*. In other words, we were actually pleased to find out that these three equations may be written directly in terms of $m_L(p)$ as follows:

$$\begin{cases} \mu(p) = \dfrac{e^{\sigma_L^2 p}}{4\pi \, [m_L(p)]^2} - p \dfrac{\sigma_L^2}{2} \\ \sigma(p) = \dfrac{e^{\frac{\sigma_L^2}{2} p}}{\sqrt{2\pi} \, m_L(p)} \\ b(p) = p - e^{\mu(p) - [\sigma(p)]^2}. \end{cases} \qquad (33)$$

The proof of Equation (33) is lengthy and was given as a special file (written in the language of the Maxima symbolic manipulator) that the reader may freely download from the web site of [13].

An important new result is now presented. The Peak-Locus Theorem Equation (33) is rewritten, not in terms of $\sigma_L$, but in terms of the four boundary parameters known as: ($ts$, $te$, $Ne$, $\delta Ne$). To this end, we must insert Equations (21) and (20) into Equation (33), producing the following result:

$$
\begin{cases}
\mu(p) = \dfrac{\left[1+\left(\frac{\delta Ne}{Ne}\right)^2\right]^{\frac{p}{te-ts}}}{4\pi\,[m_L(p)]^2} - \ln\left\{ \left[1+\left(\frac{\delta Ne}{Ne}\right)^2\right]^{\frac{p}{2(t-ts)}} \right\} \\[4mm]
\sigma(p) = \dfrac{\left[1+\left(\frac{\delta Ne}{Ne}\right)^2\right]^{\frac{p}{2(t-ts)}}}{\sqrt{2\pi}\,m_L(p)} \\[4mm]
b(p) = p - e^{\mu(p)-[\sigma(p)]^2}.
\end{cases}
\tag{34}
$$

In the particular GBM case, the mean value is Equation (22) with $\mu_{GBM} = B$, $\sigma_L = \sqrt{2B}$ and $N_0 = Ns = A$. Then, the Peak-Locus theorem Equation (33) with $ts = 0$ yields:

$$
\begin{cases}
\mu(p) = \dfrac{1}{4\pi A^2} - B\,p, \\[2mm]
\sigma = \dfrac{1}{\sqrt{2\pi}A}, \\[2mm]
b(p) = p - e^{\mu(p)-\sigma^2}.
\end{cases}
\tag{35}
$$

In this simpler form, the Peak-Locus theorem had already been published by the author in [10–12], while its most general form is Equations (33) and (34).

## 9. EvoEntropy($p$) as a Measure of Evolution

The (Shannon) Entropy of the $b$-lognormal Equation (27) is (for the proof, see [11], page 686):

$$
H(p) = \frac{1}{\ln(2)}\left[\ln\left(\sqrt{2\pi}\sigma(p)\right) + \mu(p) + \frac{1}{2}\right].
\tag{36}
$$

This is a function of the peak abscissa $p$ and is measured in bits, as in Shannon's Information Theory. By virtue of the Peak-Locus Theorem Equation (33), it becomes:

$$
H(p) = \frac{1}{\ln(2)}\left\{ \frac{e^{\sigma_L^2\,p}}{4\pi[m_L(p)]^2} - \ln(m_L(p)) + \frac{1}{2}\right\}.
\tag{37}
$$

One may also directly rewrite Equation (37) in terms of the four boundary parameters ($ts$, $te$, $Ne$, $\delta Ne$), upon inserting Equation (21) into Equation (37), with the result:

$$
H(p) = \frac{1}{\ln(2)}\left\{ \frac{\left[1+\left(\frac{\delta Ne}{Ne}\right)^2\right]^{\frac{p}{te-ts}}}{4\pi[m_L(p)]^2} - \ln(m_L(p)) + \frac{1}{2}\right\}.
\tag{38}
$$

Thus, Equation (37) and Equation (38) yield the entropy of each member of the family of $\infty^1$ $b$-lognormals (the family's parameter is $p$) peaked upon the mean value curve (1). The $b$-lognormal Entropy Equation (36) is thus the measure of the extent of evolution of the $b$-lognormal: it measures the decreasing disorganization in time of what that $b$-lognormal represents.

Entropy is thus disorganization decreasing in time. However, one would prefer to use a measure of the increasing organization in time. This is what we call the EvoEntropy of $p$:

$$
\text{EvoEntropy}(p) = -[H(p) - H(ts)].
\tag{39}
$$

The Entropy of evolution is a function that has a minus sign in front of Equation (36), thus changing the decreasing trend of the (Shannon) entropy Equation (36) into the increasing trend of this EvoEntropy Equation (39). In addition, this EvoEntropy starts at zero at the initial time $ts$, as expected.

$$
\text{EvoEntropy}(ts) = 0.
\tag{40}
$$

By virtue of Equation (37), the EvoEntropy Equation (39), invoking also the initial condition Equation (8), becomes:

$$\text{EvoEntropy}(p)\_\text{of\_the\_Lognormal\_Process\_}L(t) = \frac{1}{\ln(2)} \left\{ \frac{e^{\sigma_L^2 ts}}{4\pi Ns^2} - \frac{e^{\sigma_L^2 p}}{4\pi [m_L(p)]^2} + \ln\left(\frac{m_L(p)}{Ns}\right) \right\}. \quad (41)$$

Alternatively, this could be directly rewritten in terms of the five boundary parameters ($ts$, $Ns$, $te$, $Ne$, $\delta Ne$), upon inserting Equation (38) into Equation (39), thus finding:

$$\text{EvoEntropy}(p)\_\text{of\_the\_Lognormal\_Process\_}L(t) = \frac{1}{\ln(2)} \left\{ \frac{\left[1 + \left(\frac{\delta Ne}{Ne}\right)^2\right]^{\frac{ts}{te-ts}}}{4\pi Ns^2} - \frac{\left[1 + \left(\frac{\delta Ne}{Ne}\right)^2\right]^{\frac{p}{te-ts}}}{4\pi [m_L(p)]^2} + \ln\left(\frac{m_L(p)}{Ns}\right) \right\}. \quad (42)$$

It is worth noting that the standard deviation at the end time, $\delta Ne$, is irrelevant for the purpose of computing the simple curve of the EvoEntropy Equation (39). In fact, the latter is just a continuous curve, and not a stochastic process. Therefore, any numeric arbitrary value may be assigned to $\delta Ne$, and the EvoEntropy curve must not change. Keeping this in mind, it can be seen that the true EvoEntropy curve is obtained by "squashing" down Equation (42) into the mean value curve $m_L(t)$ and this only occurs if we let:

$$\delta Ne = 0. \quad (43)$$

Inserting Equation (43) into Equation (42), the latter can be simplified into:

$$\text{EvoEntropy}(p)\_\text{of\_the\_Lognormal\_Process\_}L(t) = \frac{1}{\ln(2)} \left\{ \frac{1}{4\pi Ns^2} - \frac{1}{4\pi [m_L(p)]^2} + \ln\left(\frac{m_L(p)}{Ns}\right) \right\} \quad (44)$$

which is the final form of the EvoEntropy curve. Equation (44) will be used in the sequel. It can now be clearly seen that the final EvoEntropy Equation (44) is made up of three terms, as follows:

(a)   The constant term

$$\frac{1}{4\pi Ns^2} \quad (45)$$

whose numeric value in the particularly important case of $Ns = 1$ is:

$$\frac{1}{4\pi} = 0.079577471545948 \quad (46)$$

that is, it approximates almost zero.

(b)   The denominator square term in Equation (44) rapidly approaches zero as $m_L(p)$ increases to infinity. In other words, this inverse-square term

$$-\frac{1}{4\pi [m_L(p)]^2}$$

may become almost negligible for large values of the time $p$.

(c)   Finally, the dominant, natural logarithmic, term, i.e., that which is the major term in this EvoEntropy Equation (45) for large values of the time $p$.

$$\ln\left(\frac{m_L(p)}{Ns}\right). \quad (48)$$

In conclusion, the EvoEntropy Equation (44) depends upon its natural logarithmic term Equation (48), and so its shape in time must be similar to the shape of a logarithm, i.e., nearly vertical at the beginning of the curve and then progressively approaching the horizontal, though never reaching it. This curve has no maxima nor minima, nor any inflexions.

## 10. Perfectly Linear EvoEntropy When the Mean Value Is Perfectly Exponential (GBM): This Is Just the Molecular Clock

In the GBM case of Equation (22) (with *t* replaced by (*t-ts*)), when the mean value is given by the exponential

$$m_{\text{GBM}}(t) = Ns\, e^{\frac{\sigma_L^2}{2}(t-ts)} = Ns\, e^{B\,(t-ts)} \tag{49}$$

the EvoEntropy Equation (44) **is exactly a linear function of the time** *p*, since the first two terms inside the braces in Equation (44) cancel each other out, as we now prove.

**Proof.** Insert Equation (49) into Equation (44) and then simplify:

$$
\begin{aligned}
&\text{EvoEntropy}(p)\_\text{of\_GBM} = \\
&= \tfrac{1}{\ln(2)}\left\{ \frac{e^{\sigma_L^2\, ts}}{4\pi Ns^2} - \frac{e^{\sigma_L^2\, p}}{4\pi\left[Ns\, e^{\frac{\sigma_L^2}{2}(p-ts)}\right]^2} + \ln\!\left(\frac{Ns\, e^{\frac{\sigma_L^2}{2}(p-ts)}}{Ns}\right) \right\} = \tfrac{1}{\ln(2)}\left\{ \frac{e^{\sigma_L^2\, ts}}{4\pi Ns^2} - \frac{e^{\sigma_L^2\, p}}{4\pi Ns^2\, e^{\sigma_L^2(p-ts)}} + \ln\!\left(e^{\frac{\sigma_L^2}{2}(p-ts)}\right) \right\} = \\
&= \tfrac{1}{\ln(2)}\left\{ \frac{e^{\sigma_L^2\, ts}}{4\pi Ns^2} - \frac{1}{4\pi Ns^2\, e^{\sigma_L^2(-ts)}} + \frac{\sigma_L^2}{2}(p-ts) \right\} = \tfrac{1}{\ln(2)}\left\{ \frac{e^{\sigma_L^2\, ts}}{4\pi Ns^2} - \frac{e^{\sigma_L^2(ts)}}{4\pi Ns^2} + \frac{\sigma_L^2}{2}(p-ts) \right\} = \\
&= \tfrac{1}{\ln(2)}\left\{ \frac{\sigma_L^2}{2}(p-ts) \right\} = \tfrac{1}{\ln(2)}\{B \cdot (p-ts)\}.
\end{aligned}
\tag{50}
$$

In other words, the GBM EvoEntropy is given by:

$$\text{GBM\_EvoEntropy}(p) = \frac{B}{\ln(2)} \cdot (p - ts). \tag{51}$$

This is a straight line in the time *p*, starting at the time *ts* of the origin of life on Earth and increasing linearly thereafter. It is measured in bits/individual and is shown in Figure 3.

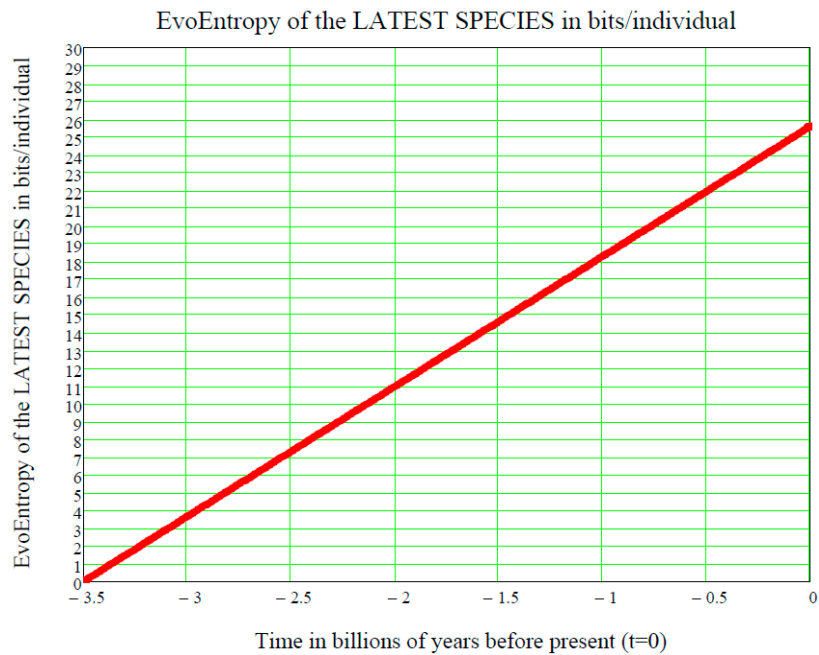

**Figure 3.** **EvoEntropy (in bits per individual) of the latest species appeared on Earth during the last 3.5 billion years if the mean value is an increasing exponential, i.e., if our lognormal stochastic process is a GBM.** This straight line shows that a Man (nowadays) is 25.575 bits more evolved than the first form of life (RNA) 3.5 billion years ago.

**This is the same linear behaviour in time as the molecular clock**, which is the technique in molecular evolution that uses fossil constraints and rates of molecular change to deduce the time in geological history when two species or other taxa diverged. The molecular data used for such calculations are usually nucleotide sequences for DNA or amino acid sequences for proteins (see [16–18]).

In conclusion, we have ascertained that the EvoEntropy in our Evo-SETI theory and the molecular clock are the same linear time function, apart from multiplicative constants (depending on the adopted units, like bits, seconds, etc.). This conclusion appears to be of key importance when assessing the stage at which a newly discovered exoplanet is in the process of its chemical evolution towards life.

## 11. Markov-Korotayev Alternative to Exponential: A Cubic Growth

Figure 3, showing the linear growth of the Evo-Entropy over the last 3.5 billion years of evolution of life on Earth, illustrates the key factor in molecular evolution and allows for an immediate quantitative estimate of how much (in bits per individuals) any two species differ from each other; this being the key to cladistics. However, after 2007, this exponential vision was shaken by the alternative "cubic vision" now outlined.

This cubic vision is detailed in the full list of papers published by Andrey Korotayev and Alexander V. Markov et al., since 2007 [1–7]. Another important publication is their mathematical paper [8] relating to the new research field entitled "Big History". In addition, a synthetic summary of the Markov-Korotayev theory of evolution appears on Wikipedia at http://en.wikipedia.org/wiki/Andrey_Korotayev, for which an adapted excerpt is seen below:

"According to the above list of published papers, in 2007–2008 the Russian scientists Alexander V. Markov and Andrey Korotayev showed that a 'hyperbolic' mathematical model can be developed to describe the macrotrends of biological evolution. These authors demonstrated that changes in biodiversity through the Phanerozoic correlate much better with the hyperbolic model (widely used in demography and macrosociology) than with the exponential and logistic models (traditionally used in population biology and extensively applied to fossil biodiversity as well). The latter models imply that changes in diversity are guided by a first-order positive feedback (more ancestors, more descendants) and/or a negative feedback arising from resource limitation. Hyperbolic model implies a second-order positive feedback. The hyperbolic pattern of the world population growth has been demonstrated by Markov and Korotayev to arise from a second-order positive feedback between the population size and the rate of technological growth. According to Markov and Korotayev, the hyperbolic character of biodiversity growth can be similarly accounted for by a feedback between the diversity and community structure complexity. They suggest that the similarity between the curves of biodiversity and human population probably comes from the fact that both are derived from the interference of the hyperbolic trend with cyclical and stochastic dynamics [1–7]."

This author was inspired by the following Figure 4 (taken from the Wikipedia site http://en.wikipedia.org/wiki/Andrey_Korotayev), showing the increase, but not monotonic increase, of the number of Genera (in thousands) during the last 542 million years of life on Earth, making up the Phanerozoic. Thus, it is postulated that the red curve in Figure 4 could be regarded as the "Cubic mean value curve" of a lognormal stochastic process, just as the exponential mean value curve is typical of GBMs.

The Cubic Equation (31) may be used to represent the red line in Figure 4, thus reconciling the Markov-Korotayev theory with our Evo-SETI theory. This is realized when considering the following numerical inputs to the Cubic Equation (31), that we derive from looking at Figure 4. The precision of these numerical inputs is relatively unimportant at this early stage of matching the two theories (this one and the Markov-Korotayev's), as we are just aiming for a "proof of concept", and better numeric approximations might follow in the future.

$$
\begin{cases}
ts = -530 \\
Ns = 1 \\
te = 0 \\
Ne = 4000.
\end{cases}
\tag{52}
$$

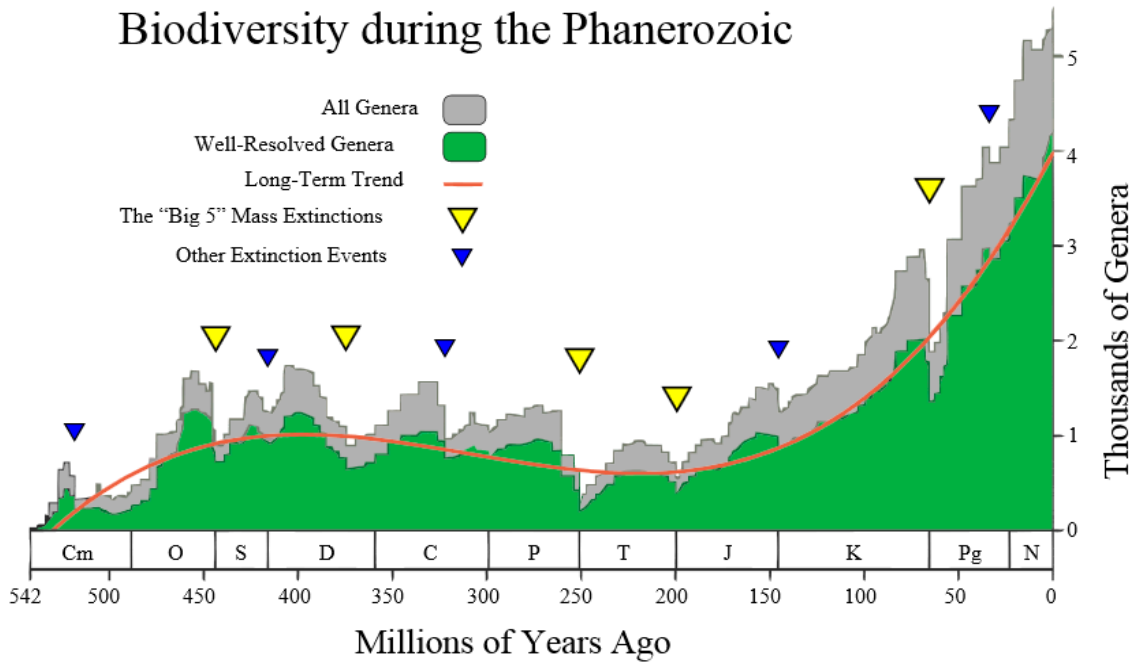

**Figure 4.** According to Markov and Korotayev, during the Phanerozoic, the biodiversity shows a steady, but not monotonic, increase from near zero to several thousands of genera.

In other words, the first two equations of Equation (52) mean that the first of the genera appeared on Earth about 530 million years ago, i.e., the number of genera on Earth was zero before 530 million years ago. In addition, the last two equations of Equation (52) mean that, at the present time $t = 0$, the number of genera on Earth is approximately 4000, noting that a standard deviation of about $\pm 1000$ affects the average value of 4000. This is shown in Figure 4 by the grey stochastic process referred to as all genera. It is re-phrased mathematically by assigning the fifth numeric input:

$$
\delta Ne = 1000.
\tag{53}
$$

Then, as a consequence of the five numeric boundary inputs ($ts$, $Ns$, $te$, $Ne$, $\delta Ne$), plus the standard deviation $\sigma$ on the current value of genera, Equation (19) yields the numeric value of the positive parameter $\sigma$:

$$
\sigma = \sqrt{\frac{\ln\left[1 + \left(\frac{\delta Ne}{Ne}\right)^2\right]}{te - ts}} = 0.011.
\tag{54}
$$

Having thus assigned numerical values to the first five conditions, only the conditions on the two abscissae of the Cubic maximum and minimum, respectively, tone to be assigned. Figure 4 establishes them (in millions of years ago) as:

$$
\begin{cases}
t_{\text{Max}} = -400 \\
t_{\text{min}} = -220.
\end{cases}
\tag{55}
$$

Finally, inserting these seven numeric inputs into the Cubic Equation (31), as well as into both of the equations of Equation (15) of the upper and lower standard deviation curves, the final plot shown in Figure 5 is produced.

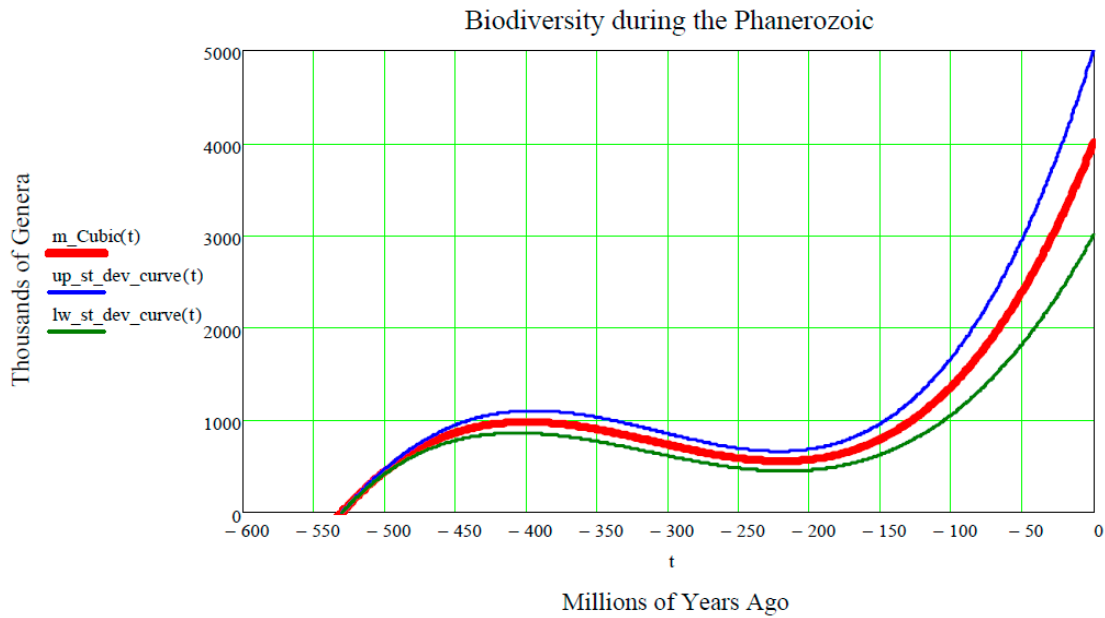

**Figure 5.** The Cubic mean value curve (thick red solid curve) ± the two standard deviation curves (thin solid blue and green curve, respectively) provide more mathematical information than Figure 4. One is now able to view the two standard deviation curves of the lognormal stochastic process, Equations (11) and (12), that are completely missing in the Markov-Korotayev theory and in their plot shown in Figure 4. This author claims that his Cubic mathematical theory of the Lognormal stochastic process $L(t)$ is a more profound mathematization than the Markov-Korotayev theory of Evolution, since it is stochastic, rather than simply deterministic.

## 12. EvoEntropy of the Markov-Korotayev Cubic Growth

What is the EvoEntropy Equation (44) of the Markov-Korotayev Cubic growth Equation (31)?

To answer this question, Equation (31) needs to be inserted into Equation (44) and the resulting equation can then be plotted:

$$\text{Cubic\_EvoEntropy}(t) = \frac{1}{\ln(2)} \cdot \left\{ \frac{1}{4\pi Ns^2} - \frac{1}{4\pi [m_{\text{Cubic}}(t)]^2} + \ln\left(\frac{m_{\text{Cubic}}(t)}{Ns}\right) \right\}. \tag{56}$$

The plot of this function of $t$ is shown in Figure 6.

## 13. Comparing the EvoEntropy of the Markov-Korotayev Cubic Growth, to the Hypothetical (1) Linear and (2) Parabolic Growth

It is a good idea to consider two more types of growth in the Phanerozoic:

(1)　The LINEAR (= straight line) growth, given by the mean value of Equation (29)
(2)　The PARABOLIC (= quadratic) growth, given by the mean value of Equation (30).

These can be compared with the CUBIC growth Equation (31) typical for the Markov-Korotayev model.

The results of this comparison are shown in the two diagrams (upper one and lower one) in Figure 7.

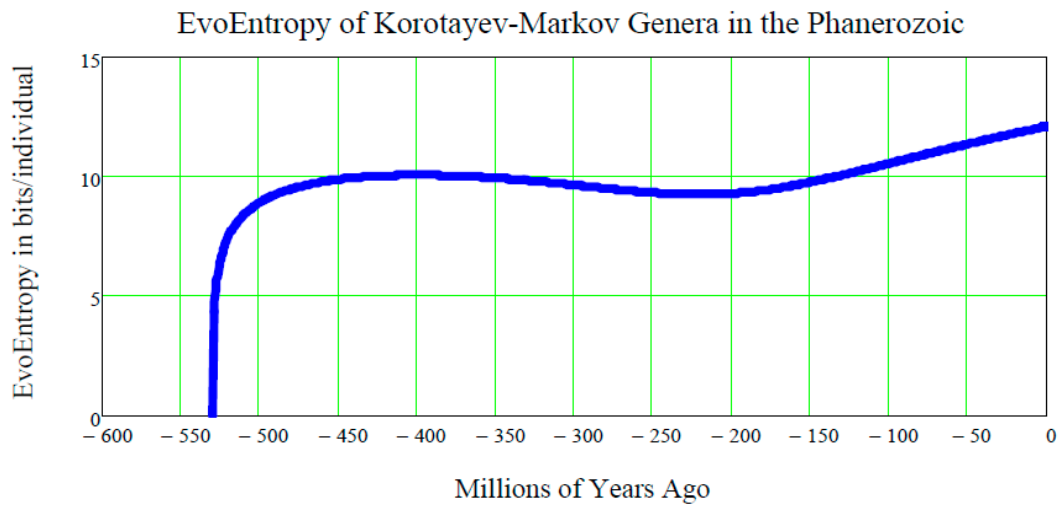

**Figure 6. The EvoEntropy Equation (44) of the Markov-Korotayev Cubic mean value Equation (31) of our lognormal stochastic process $L(t)$ applies to the growing number of Genera during the Phanerozoic.** Starting with the left part of the curve, one immediately notices that, in a few million years around the Cambrian Explosion of 542 million years ago, the EvoEntropy had an *almost vertical growth*, from the initial value of zero, to the value of approximately 10 bits per individual. These were the few million years when the *bilateral symmetry* became the dominant trait of all primitive creatures inhabiting the Earth during the Cambrian Explosion. Following this, for the next 300 million years, the EvoEntropy did not significantly change. This represents a period when bilaterally-symmetric living creatures, e.g., reptiles, birds, and very early mammals, etc., underwent little or no change in their body structure (roughly up to 310 million years ago). Subsequently, after the "mother" of all mass extinctions at the end of the Paleozoic (about 250 million years ago), the EvoEntropy started growing again in mammals. Today, according to the Markov-Korotayev model, the EvoEntropy is about 12.074 bits/individual for humans, i.e., much less than the 25.575 bits/individual predicted by the GBM exponential growth shown in Figure 3. Therefore, the question is: which model is correct?

For the sake of simplicity, we omit all detailed mathematical calculations and confine ourselves to writing down the equation of the:

(1)    LINEAR EvoEntropy:

$$\text{STRAIGHT\_LINE\_EvoEntropy}(t) = \frac{1}{\ln(2)}\left\{\frac{1}{4\pi Ns^2} - \frac{1}{4\pi\left[m_{\text{straight\_line}}(t)\right]^2} + \ln\left(\frac{m_{\text{straight\_line}}(t)}{Ns}\right)\right\}. \tag{57}$$

(2)    PARABOLIC (quadratic) EvoEntropy:

$$\text{PARABOLA\_EvoEntropy}(t) = \frac{1}{\ln(2)}\left\{\frac{1}{4\pi Ns^2} - \frac{1}{4\pi\left[m_{\text{parabola}}(t)\right]^2} + \ln\left(\frac{m_{\text{parabola}}(t)}{Ns}\right)\right\}. \tag{58}$$

(3)    CUBIC (MARKOV-KOROTAYEV) EVOENTROPY, i.e., Equation (56).

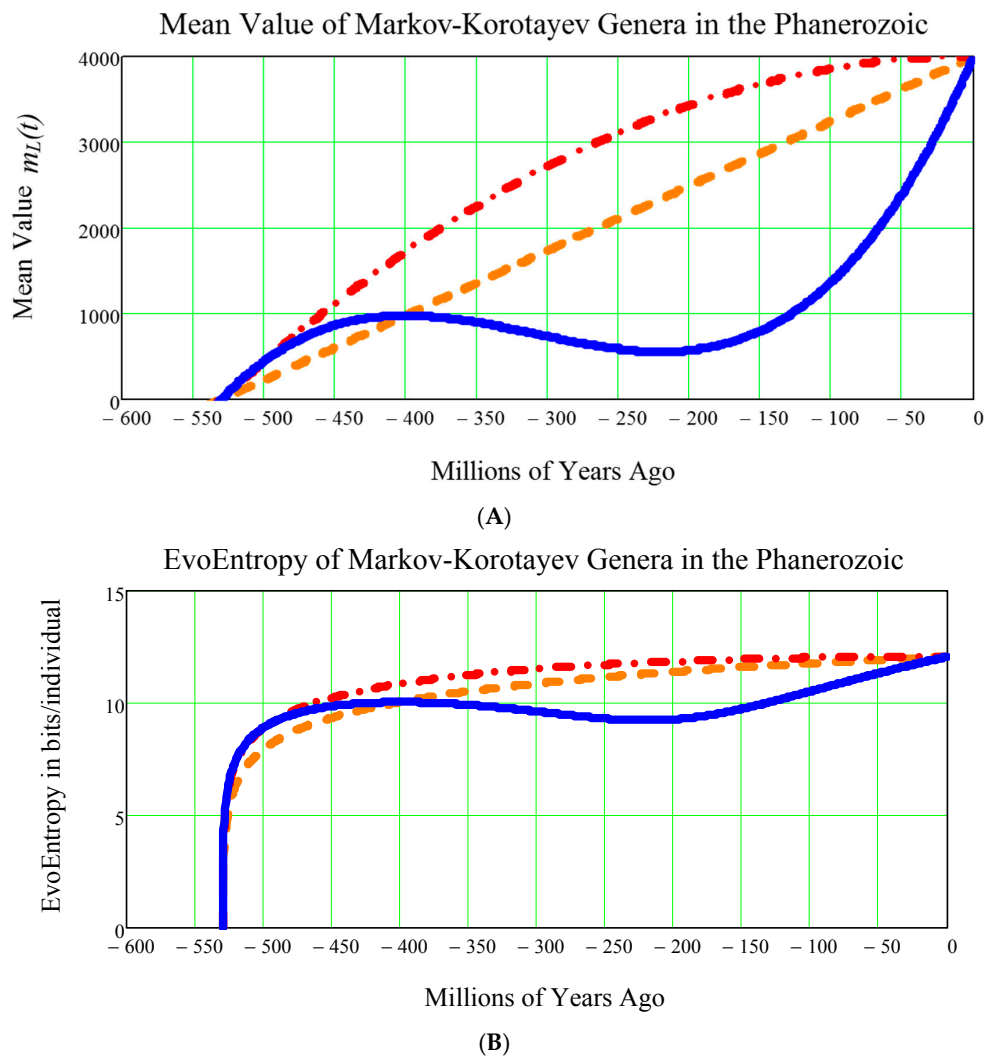

**(A)**

**(B)**

**Figure 7.** Comparing the mean value $m_L(t)$ **(A)** and the EvoEntropy($t$) **(B)** in the event of growth with the CUBIC mean value of Equation (31) (blue solid curve), with the LINEAR Equation (29) (dash-dash orange curve), or with the PARABOLIC Equation (30) (dash-dot red curve). It can be seen that, for all these three curves, starting with the left part of the curve, in a few million years around the Cambrian Explosion of 542 million years ago, the EvoEntropy had an almost vertical growth from the initial value of zero to the value of approximately 10 bits per individual. Again, as is seen in Figure 6, these were the few million years where the bilateral symmetry became the dominant trait of all primitive creatures inhabiting the Earth during the Cambrian Explosion.

## 14. Conclusions

The evolution of life on Earth over the last 3.5 to 4 billion years has barely been demonstrated in a mathematical form. Since 2012, I have attempted to rectify this deficiency by resorting to lognormal probability distributions in time, starting each at a different time instant $b$ (birth), called $b$-lognormals [9–14,19]. My discovery of the Peak-Locus Theorem, which is valid for any enveloping mean value (and not just the exponential one (GBM), for the general proof see [16], in particular supplementary materials over there), has made it possible for the use of the Shannon Entropy of Information Theory as the correct mathematical tool for measuring the evolution of life in bits/individual.

In conclusion, the processes which occurred on Earth during the past 4 billion years can now be summarized by statistical equations, noting that this only relates to the evolution of life on Earth, and not on other exoplanets. The extending of this Evo-SETI theory to life on other exoplanets will only be possible when SETI, the current scientific search for extra-terrestrial intelligence, achieves the first contact between humans and an alien civilization.

**Supplementary Materials:** They are available online at www.mdpi.com/2075-1729/7/2/18/s1.

**Conflicts of Interest:** The author declares no conflict of interest.